# Primary damage production in the presence of extended defects and growth of vacancy-type dislocation loops in hcp zirconium

Cong Dai, Fei Long, Peyman Saidi, Laurent Karim Béland*, Zhongwen Yao, Mark R. Daymond*

Department of Mechanical and Materials Engineering, Queen's University, Kingston, ON K7L 3N6, Canada

* Corresponding authors; e-mail addresses: laurent.beland@queensu.ca; mark.daymond@queensu.ca

**Abstract**

Production rates in long-term predictive radiation damage accumulation models are generally considered independent of the material's microstructure for reactor components. In this study, the effect of pre-existing microstructural elements on primary damage production in α-Zr – and vice-versa—is assessed by molecular dynamics (MD) simulations. $a$-type dislocation loops, $c$-component dislocation loops and a tilt grain boundary (GB) were considered. Primary damage production is reduced in the presence of all these microstructural elements, and clustering behavior is dependent on the microstructure. Collision cascades do not cause $a$-type loop growth or shrinkage, but they cause $c$-component loop shrinkage. Cascades in the presence of the GBs produce more vacancies than interstitials. This result, as well as other theoretical, MD and experimental evidence, confirm that vacancy loops will grow in the vacancy supersaturated environment near GBs. Distinct temperature-dependent growth regimes are identified. Also, MD reveals cascade-induced events where $a$-type vacancy loops are absorbed by GBs. Fe segregation at the loops inhibits this cascade-induced absorption.

# 1 Introduction

Long-term exposure to radiation causes structural damage and limits the lifetimes of nuclear reactor components, many made from Zr alloys. Irradiation-induced point defects rearrange to form point-defect clusters, dislocation loops and voids [1]. In particular, in Zr alloys $c$-component loops are believed to have a significant impact on irradiation growth and irradiation creep as well as the degradation of mechanical behavior [2]. In addition, the formation and distribution of irradiation-induced point defects are determined by the material's prior microstructure [3]. For instance, early experiments by Griffiths et al. [4] on neutron-irradiated Zr show that interstitial loops were estimated to make up about 25% of the total dislocation loops observed near grain boundaries (GBs), for a 2~3 μm grain size.

Irradiation damage of Zr alloys has been extensively studied using transmission electron microscopy (TEM) [5], X-ray analysis [6] and positron annihilation spectroscopy [7]. However, these experimental instruments have limitations in providing detailed information in regards to the nano-size defects that form during pico-second collision cascades. Thus, different computational and modeling techniques such as molecular dynamics (MD), kinetic Monte Carlo (kMC) and rate theory have been applied to various size- and time-scale problems.

Given the small MD timesteps—typically $\sim 1 \times 10^{-15}$s, and as small as $1 \times 10^{-18}$s in collision cascade simulations—the evolution of radiation-induced damage that can be taken into account by MD is of the order of a few tens of nanoseconds. There are three main stages to primary damage production that take place over a few picoseconds: a supersonic phase, a sonic phase and a thermally enhanced recovery phase [8]. MD is well suited to making physically valid predictions over these timescales, within the limits of the choice of interatomic potentials (see Refs. [9-13] for discussions about designing potentials for cascade simulations). MD can also uncover atomistic phenomena that take place on timescales of a few nanoseconds, revealing relevant mechanisms that should be included in kMC and rate-theory models, such as 1-D and 3-D diffusion of defects [14-16], or interaction of point defects, defect clusters and extended defects such as dislocation loops and line dislocations [17-19].

To calculate damage accumulation and evolution over longer timescales, the results of these MD simulations--damage production rates and clustering behavior—are typically used as inputs in kMC and rate-theory based models. For example, Arévalo et al. [20,21] used a database of displacement cascades from MD to study the effect of temperature on the accumulation of damage in pure Zr using kMC. Likewise, using kMC calculations, the anisotropic diffusion of point defects in Zr was analyzed by Fan et al. [22]. For even longer-time treatments, rate theory models are used, determined by the coupled reaction-diffusion equations [23-25]. Woo et al. [26] proposed a rate theory that considers anisotropic diffusion of point defects. They found that the difference in diffusional anisotropy (DAD) between vacancies and interstitials could cause a large bias in their reaction rates with sinks, even if the dislocation structure was isotropic. The DAD between vacancies and interstitials was justified by molecular dynamics simulations based on empirical potentials. Later, Samolyuk et al. [26] showed, using atomistic simulations based on electronic structure calculations, that DAD was much lower than that predicted using empirical potentials. The *ab initio* based DAD alone cannot explain radiation-induced growth of Zr below 750K. Interestingly, including one-dimensional diffusion of cascade-induced SIA-clusters in the rate theory—which were revealed by MD—can lead to reasonable predictions of radiation-induced growth [27,28]. A key takeaway here is that the predictions—and physical validity—of rate theories are largely determined by the mechanisms and rates that they include, which are often provided by atomistic simulations. Recently, Payant [29] used cluster dynamics models to predict the microstructure evolution as a function of dose and temperature, and predicted significantly higher loop densities than those experimentally observed. They emphasized the importance of MD simulations in providing a database on the effects of collision cascades with pre-existing defects.

Such long timescale models typically neglect the effect of pre-existing microstructure on primary damage production rates. This is a good first-order approximation, but it may break down as the densities of extended defects increase as damage accumulates. It may also break down in the case of nano-grained materials, which are designed to have a very high density of grain boundaries.

The effect of pre-existing microstructure on primary damage production has been assessed by a number of previous MD-based studies. Ludy et al. [30] studied collision cascades induced by primary knock-on atoms (PKAs) with kinetic energies of up to 2.5 keV near high-angle GBs in Cu, and found that GBs easily absorb SIAs while more residual vacancies remain in the bulk. This suggests that high-angle GBs in Cu capture SIAs *in-cascade*. The cascade sink preference of GBs in α-Zr was also studied [31,32]. The cascade sink is defined as the asymmetry between interstitial-type and vacancy-type primary damage production when cascades occur in non-pristine materials. Performing cascade simulations in the vicinity of five different GB structures, Hatami et al. [32], observed that GBs in α-Zr are not necessarily biased toward interstitials, and can preferentially absorb vacancies. Hatami et al. considered PKAs with initial kinetic energy less than 9 keV, and neglected the effect of PKA direction. Jin et al. [33] analyzed radiation damage evolution in Cu bicrystals by simulating overlapping cascades, and they reported a mechanism for annihilation of defect clusters during irradiation. Jin et al. also discussed how the short simulation time between collision cascades yields a significantly higher effective dose rate than that observed in experiments. This time-scale problem limits MD in the investigation of diffusion-based microstructural evolution.

As mentioned above, realistic MD data in regards to defect production are valuable input parameters for kMC and rate theory models [34]. MD can help us understand in detail the accumulation of damage in irradiated materials, e.g., intra-cascade recombination, the creation of primary clusters and cascade-induced growth or shrinkage of pre-existing extended defects. Augmenting cascade-induced event rates databases will improve model-based predictions, and provide a more comprehensive view of radiation damage [35]. Likewise, simulating the kinetics of defects over timescales of nanoseconds may reveal mechanisms that the higher-scale models need to take into account. Our study has thus focused on the following three themes. First, primary cascade production in the presence of dislocation loops. Second, the growth of dislocation loops near GBs. Third, the stability of the dislocation loop near GBs and the effect of alloying on this stability.

MD is used to simulate collision cascades in α-Zr that overlap with extended defects: *a*-type or *c*-component dislocation loops, and a tilt GB. A mixture of theoretical arguments, MD simulations, and experimental evidence—*in situ* transmission electlron micrographs (TEM) of Zircaloy-2 irradiated with 3 MeV proton beam—are combined to show that vacancy dislocation loops near GBs grow in a vacancy-supersaturated environment. MD simulations of *a*-type and *c*-component loop growth in this environment are reported. The stability of these loops near a grain boundary under irradiation is assessed by MD. The effect of Fe solutes on this stability is assessed by a hybrid molecular dynamics/Monte Carlo (MD/MC) scheme.

## 2 Methods

### 2.1 Simulation description

The MD was simulated using the LAMMPS (Large-scale Atomic/Molecular Massively Parallel Simulator) [36]. There are a few [37-39] Zr potentials available. As reported in [40], the MA07 [39] potential (#2) provides a better description of vacancy binding energies than the MA07 [39] potential (#3). However, the latter (#3) provides a better description of the stacking faults on prismatic, basal and pyramidal slip systems [40], which largely controls the energetic of extended defects such as dislocation loops. We chose the latter one (#3). Also, it should be noted that the short-range interactions predicted by this potential are stiffened, which is key when predicting primary damage production [9,10,41]. This potential has been used in a number of recent radiation damage simulation studies [42-46]. An asymmetric $(1\bar{1}00)/(11\bar{2}0)\langle 0002\rangle(\theta=30°)$ tilt grain boundary was created. Its boundary plane orientations are the $(1\bar{1}00)$ and $(11\bar{2}0)$ surfaces of the two grains. This GB is illustrated in Figure 1. The GB structure was constructed using a conjugate gradient energy minimization [47] in conjunction with an atom deletion criterion [48]. Periodic boundary conditions are imposed along all directions (*X, Y* and *Z* axis). Typical samples contain 4 million atoms with dimensions $L_x$=44.5 nm, $L_y$=56.0 nm and $L_z$=20.6 nm.

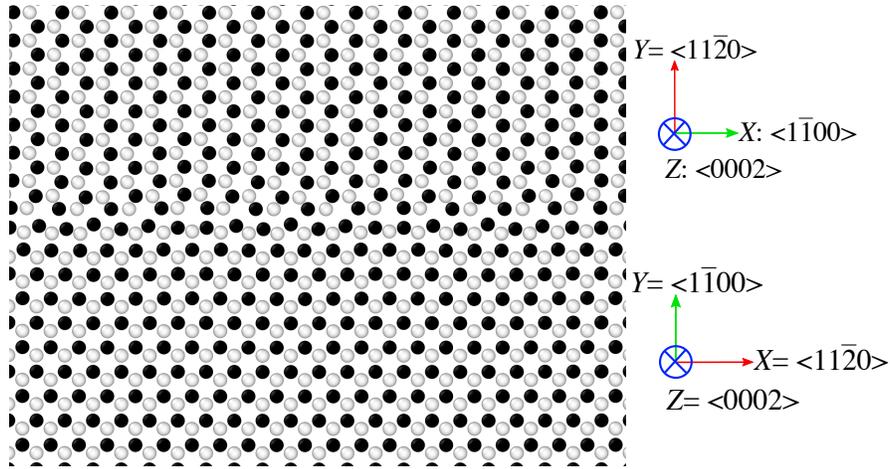

**Figure 1. The asymmetric tilt grain boundary $(1\bar{1}00)/(11\bar{2}0)\langle0002\rangle(\theta=30°)$ is viewed along the $\langle0002\rangle$ tilt axis; atoms on consecutive {0002} planes are shown as black and white. The crystal orientations are shown in the left-hand side for each for grain.**

Cascade simulations were carried out at 573 K, which is a common environmental temperature during irradiation in nuclear reactors [1]. Within the time of collision cascade simulations using MD, only short-range diffusion is considered. The purpose of the cascade simulations in this study is to investigate primary damage production in the presence of different pre-existing microstructures. In this context, temperature-effects should be relatively minor. For this reason, only one cascade simulation temperature was considered. Long-range diffusion—a process that crucially depends on temperature—is largely ignored. Moreover, recent cascade simulations by Nordlund et al. [13] suggests that the effect of temperature on primary damage production is minor, an observation consistent with several previous studies [34,49,50]. The energy of a PKA was set to be 50 keV, and the directions of the PKA were randomly selected. Ions and electrons both significantly contribute to stopping of 50 keV ions. In this study, we solely considered ion stopping. In the future, it would be interesting to consider the influence of electron stopping and electron-phonon coupling on cascade-microstructure interactions, e.g., using a two-temperature model [51,52]. It should be noted, however, that the present article focuses on differences between cascades in the bulk and cascades in the presence of pre-existing microstructure. We expect that the absolute level of primary damage production is likely to be affected by electron stopping and electron-phonon coupling to a much larger extent than microstructure-induced differences in primary damage production. This assumption and ensuing computational strategy are commonly used to simulate radiation damage in this energy range [53-55]. Also, future two-temperature model-based work should likely take into account the effect of pre-existing microstructure on the coupling between the atomic and electronic subsystems. For instance, the approach suggested by Tamm et al. [56,57] could prove fruitful in such an endeavour.

At least 10 independent cascade simulations were employed to obtain statistics. The cascade simulations were performed with a constant number of atoms, volume and energy (NVE ensemble) with a variable timestep.

The Zr-Fe interactions are described by the EAM developed by Saidi et al. [58]. Their model can accurately reproduce the formation energy and lattice parameter of all five stable and metastable intermetallic compounds of the Zr-Fe phase diagram. To simulate alloying segregation, the variance-constrained semi-grand-canonical (VC-SGC) ensemble [59] was employed, which combines MD for structural relaxation, and MC to push the atomic chemical configuration toward equilibrium. The fraction of swapped atoms is 0.2 and the MD runs between the MC swaps comprised 100 integration steps of 1 fs each. The chemical composition is controlled by the chemical potential difference between the two species and the target concentration. Images of atomic configurations were produced with the Open Visualization Tool (OVITO) [60]. A Wigner-Seitz cell method [61-63] was used to identify the type of point defects, where atom positions are analyzed with respect to a geometric structure with perfect lattice atom positions.

## 2.2  Experiments

Annealed alloy Zircaloy-2 was irradiated by a 3MeV proton beam at 180°C and at a flux of $3.26\times10^{18}$ ion/cm$^2$ (~ 0.11 displacement per atom) on a 15mm×15mm area with the Tandem accelerator at the Reactor Material Testing Laboratory (RMTL), Queen's University. TEM foils were prepared by back-polishing and subsequent thinning using a solution of 10% perchloric acid and 90% methanol at -40°C using a Struers Tenupol-5. TEM imaging was carried out on an FEI Tecnai Osiris Scanning/TEM (S/TEM). In order to observe the growth behavior of the irradiation-induced loops and the effect of grain boundary as a defect sink, *in situ* TEM annealing experiments were carried out with one irradiated sample loaded in a Gatan 625 double tilt heating stage at 673 K. Bright field (BF) images were taken from three areas close to the grain boundary in one grain with the same reflection prior, during, and at the end of heating, so that the change of a given group of defects could be monitored.

## 3 Results and Discussions

### 3.1 MD simulations: the primary cascade with different extended defects

The interactions of cascades with a perfect crystal, $a$-type loops, $c$-component loops and a tilt GB are statistically studied and reported in Table 1. An illustration of residual defects identified using the Wigner-Seitz cell method is provided in the Supplementary Materials [64]. Cascades produce fewer residual interstitial-type defects $N_I$ and residual vacancy-type defects $N_V$ in the simulation boxes containing a pre-existing defect than in the perfect crystal. Vacancy and interstitial production is balanced in the presence of $a$-type loops, and is imbalanced in the presence of $c$-component loops and a tilt GB. As described in Table 1, the value of $N_V$ is almost half that of $N_I$ for $c$-component interstitial loops, while the value of $N_I$ is almost half of $N_V$ for $c$-component vacancy loops and a tilt GB. This imbalanced cascade production is associated with a shrinkage of $c$-component loops and vacancy supersaturation near GBs.

Clustering of cascade-induced defects is characterized by calculating the number of interstitial clusters $N_C^I$ and the number of vacancy clusters $N_C^V$, as well as the proportion of interstitial/vacancy defects in clusters $P_C^I$ and $P_C^V$. The number of clusters does not seem altered by $a$-type loops, but it is decreased by $c$-component loops and a tilt GB. In particular, $N_C^I$ near a tilt GB is less than half that of a perfect crystal, which is consistent with the comparison between $N_I$ in the two cases. Moreover, vacancies have a lesser propensity to cluster in the presence of *a*-type and *c*-component loops, but have the same propensity to cluster in the presence of a tilt GB as in a perfect crystal.

The production rate of interstitials and vacancies is generally assumed to be identical in rate theory [23,25], and the fraction of interstitial clusters and vacancy clusters is only analyzed in a perfect crystal [28]. Also, cascade-induced shrinkage of *c*-component loops is neglected. The results presented in Table 1 could be used to modify rate theory and kMC models.

Table 1. The number of residual interstitials ($N_I$), the number of residual vacancies ($N_V$), the number of interstitial clusters ($N_C^I$), the number of vacancy clusters ($N_C^V$), the proportion of interstitial defects contained in clusters ($P_C^I$) and the proportion of vacancy defects contained in clusters ($P_C^V$) after cascade simulations with 50 keV PKA energies at 573 K. The simulation boxes contained either a perfect crystal, an *a*-type interstitial or vacancy loop, an *c*-component interstitial or vacancy loop, or a tilt GB.

|  | Pristine | *a*-type interstitial loop | *a*-type vacancy loop | *c*-component interstitial loop | *c*-component vacancy loop | Tilt GB |
|---|---|---|---|---|---|---|
| $N_I$ | 78 (6.1) | 65 (5.2) | 54 (5.0) | 67 (8.6) | 28 (5.2) | 37 (6.0) |
| $N_V$ | 78 (6.1) | 63 (5.4) | 63 (4.5) | 34 (9.1) | 41 (8.9) | 80 (11.5) |
| $N_C^I$ | 7 (1.3) | 6 (0.5) | 6 (0.5) | 6 (1.1) | 4 (0.9) | 3 (0.7) |
| $N_C^V$ | 4 (0.8) | 4 (0.4) | 4 (0.3) | 3 (0.6) | 4 (0.5) | 4 (0.5) |
| $P_C^I$ | 66.83% (5.23%) | 66.62% (2.35%) | 58.88% (2.9%) | 79.64% (4.39) | 70.21% (2.73%) | 69.69% (5.08%) |
| $P_C^V$ | 81.8% (4.14%) | 63.95% (3.61%) | 60.83% (3.67%) | 69.6% (4.91%) | 74.02% (6.34%) | 82.68% (3.65%) |

### 3.2 Evidence for vacancy supersaturation near grain boundaries

In the previous section, it was observed that GBs promote vacancy supersaturation by acting as a cascade sink. However, this result might depend on the grain boundary orientation [32,65,66]. Here, by combining classical diffusion sink arguments, MD simulations of dislocation loop stability and experimental evidence, vacancy supersaturation near GB is established in a definitive manner.

In a first approximation, a GB is a perfect sink for all point defects and defect clusters during the long-time annealing stage. It is well-known that diffusion coefficients for SIAs and their clusters are much higher than those of vacancies and their clusters in α-Zr, respectively [67,68]. If reactions between defects are neglected, a simple Fickian diffusion model will predict a vacancy surplus near the GBs.

#### 3.2.1 The stability of $a$-type dislocation loops near a tilt GB

GBs might affect the damage accumulation and evolution not only by serving as a cascade sink and a diffusion sink, but also by absorbing certain types of loops. We studied the thermal stability of $a$-type dislocation loops near a GB. An interstitial loop was created 10 nm below the tilted GB, and its initial habit plane is $(10\bar{1}0)$ as displayed in Figure 2 (a). After relaxation at 573 K for 101 ps, this loop began to tilt, and it had a chair-shape structure after relaxation for 330 ps (see Figure 2 (c)). At 340 ps, half of the loop was absorbed by the GB and it was then completely absorbed as illustrated in Figure 2 (e). The potential energy profile as a function of time is provided in Figure 2 (f), which decreases as the loop approaches the GB. An $a$-type vacancy loop was also created below the GB at the same separation distance. This $a$-type vacancy loop in the plane $(10\bar{1}0)$ was stable during the relaxation at 573 K. Thus, $a$-type vacancy loops are thermally more stable than the $a$-type interstitial loops near the GB.

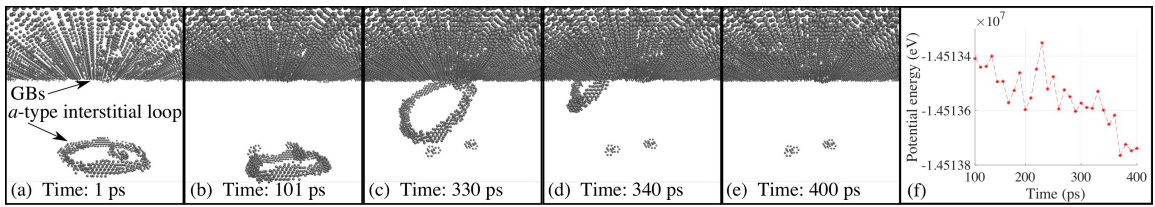

**Figure 2. A thermal relaxation simulation illustrating the absorption of an $a$-type interstitial dislocation loop by the GB at 573 K. The clock starts when the $a$-type interstitial loop is formed. Only displaced atoms are shown based on common neighbor analysis. (f) is the potential energy profile of the system as a function of time.**

### 3.3 *In situ* TEM observation: identifying the type of loops near the GBs

The previous sections of this article bring forward three theoretical results pertaining to defects near GBs. First, MD revealed that cascades generate more vacancies than interstitial near a GB. Second, we quickly highlighted that—given that vacancies are less mobile than interstitials—simple Fickian arguments predict that, near a sink such as a GB, the concentration of vacancies should be higher than the concentration of self-interstitials. Third, MD revealed that $a$-type vacancy loops are more stable than $a$-type interstitial loops near a GB. These three results suggest that dislocation loops near grain boundaries will mainly be vacancy-type. In this section, we present experiments that were carried out to analyze the type of dislocation loops near the GBs.

Most of the loops generated by 3MeV proton irradiation are smaller than 10nm. Their vacancy or interstitial nature cannot be determined because of their small size. The loops were annealed for three hours at 673 K to increase their size due to absorption of point defects (mostly vacancy) and subsequent coarsening, and hence let us identify their vacancy/interstitial nature using BF images. According to the inside/outside contrast method proposed by Maher and Eyre [69], and Föll and Wilkens [70], one needs to determine the loop Burgers vector $b$, habit plane normal $n$, the angle between the diffraction vector $g$ and the habit plane normal $n$, and also the sign of the Bragg deviation parameter $s_g$. The BF images in Figure 3 were taken with $(10\bar{1}\bar{1})$ and $(\bar{1}011)$ reflection with deviation parameter $s_g>0$ along the $[11\bar{2}3]$ zone at the same area after *in situ* annealing, as shown in Figure 3 (a) and (b) respectively. As indicated by red arrows, almost all the loops have shown outside contrast with the $(\bar{1}011)$ reflection. Given that loops with $\pm a/3[1\bar{2}10]$ Burgers' vector are invisible with this reflection, the loops shown in Figure 3 have either $\pm a/3[11\bar{2}0]$ or $\pm a/3[\bar{2}110]$ Burgers vector. It is known that loops show outside contrast when $(g\cdot b)\, s_g > 0$ [71]. The Burgers' vectors of loops that gives outside contrast are determined to be $a/3[\bar{1}\bar{1}20]$, or $a/3[\bar{2}110]$. Since $(B\cdot b)$ <0 ($B$ is the electron beam direction that is nearly parallel to the $[11\bar{2}3]$ zone axis) for both determined Burgers' vectors, this implies that all the loops showing outside contrast are vacancy type. Vacancy loops near a grain boundary grow at a higher rate than those further away. As indicated by green arrows, only a few small interstitial loops can be determined in Figure 3. Note that the observation was conducted *in-situ* with video recording; no dislocation loop disappeared during annealing. In addition, a quantitative analysis provided in the Supplementary document [64] suggests that almost all small loops present prior to annealing are also vacancy type in nature. Therefore, the overall numerical distribution of interstitial and vacancy loops determined after annealing is believed to be representative of the as-irradiated microstructure. These results are in good agreement with both Holt's study [72] and that of Griffiths et al. [4], which showed a depletion of interstitial loops near the GBs. That is, a dominance of vacancy loops as found near GBs by TEM experiments is supported by the MD results reported above which show that $a$-type vacancy loops are more stable than $a$-type interstitial loops near the GBs.

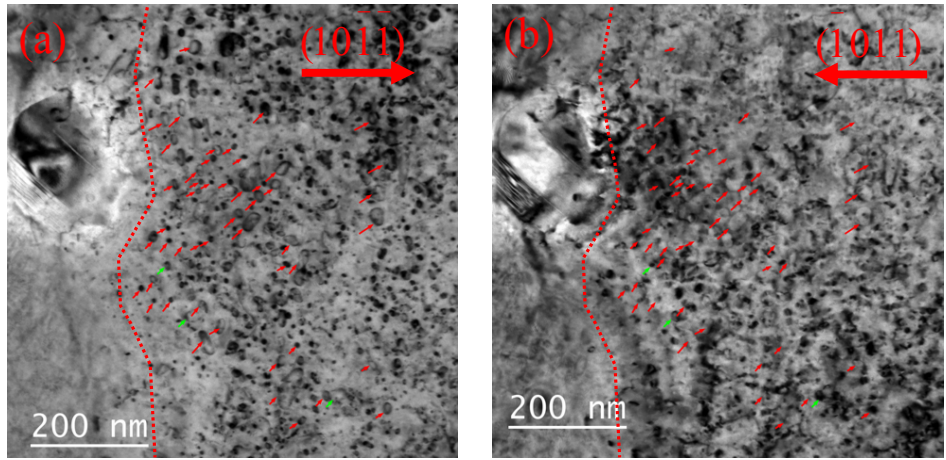

**Figure 3. Bright Field images taken with $(10\bar{1}\bar{1})$ and $(\bar{1}011)$ reflection along the $[11\bar{2}3]$ zone axis from a sample annealed at 673 K for three hours. Red arrows indicating vacancy loops near grain boundary have larger size distribution.**

### 3.4 The growth of vacancy loops under locally high concentration of vacancies

As shown in the previous section, supersaturation of vacancies near GBs leads to the growth of vacancy loops. MD was performed to simulate the loop growth in a supersaturated vacancy environment. In the vicinity of either one $a$-type or one $c$-component vacancy loop, 0.5% atoms (~287,400 atoms in total) were randomly deleted to create a locally high-concentration of vacancies, which is much higher than the equilibrium vacancy concentration [72], but roughly representative of typical point defect concentrations observed in irradiated environments [73,74].

Three annealing temperatures (600 K, 800 K, and 1000 K) were employed. The area of dislocation loops $A(t)$ was measured at each timestep, and the difference of the area regarding to its initial size ($A_0$) is defined as $\Delta A = A(t) - A_0$. Figure 4 (a) shows $\Delta A$ as a function of time for $a$-type and $c$-component vacancy loops. The growth rate of vacancy loops increases as temperature increases. The growth rate of $a$-type loops is greater than that of $c$-component vacancy loops at low temperatures. At 600 K, both vacancy loops show a very fast growth rate for the first 1 ns, and a slower rate afterwards. Visual inspection of loop growth at 600 K indicates that the loop can easily absorb nearby vacancies in the beginning of the simulation, which leads to a faster growth rate. Once the nearby vacancies are captured, the growth rate slows. At 1000 K, the growth rates for both loops are roughly constant with time over the period investigated.

Vacancy clusters formed throughout the simulation boxes. In addition, $a$-type vacancy loops can locally glide and tilt, while $c$-component vacancy loops remain fixed in the basal plane. This suggests that $a$-type vacancy loops may be more readily absorbed by the GBs as loops tilt and glide.

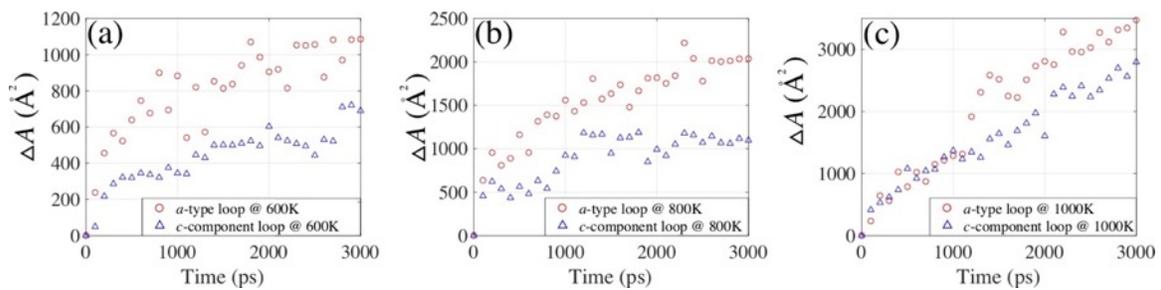

**Figure 4. Loop area differences as a function of annealing time at (a) 600 K, (b) 800 K and (c) 1000 K.**

In addition to growth rates, atomic diffusion was monitored. Mean square displacement (MSD) in different directions are summed over all the atoms in the system, and illustrated in Figure 5. The time-dependent MSD shows three main features: (i) the MSD at 600 K and 800 K exhibits a supralinear trend; (ii) at 600 K, the MSD for a system having an *a*-type (*c*-component) vacancy loop is highest in the *y (z)* direction; and (iii) the MSD traces in all directions are similar at 1000 K, but the slop of the MSD slightly decreases after 1.5 ns.

The supralinear MSD trend at lower temperatures might be due to interaction between the loops and the vacancy clusters. Visual inspection indicates that the vacancy absorption occurs at the edge of the loop. Let us therefore further examine the MSD results at a low temperature in Figure 5. The normal direction of the *a*-type vacancy loop's plane at 600 K is parallel to the *y* direction, which is the same direction for the highest MSD in Figure 5 (a). At the same temperature, the habit plane of *c*-component vacancy loops is the basal plane, and its normal direction is also the same direction for the highest MSD in Figure 5 (d). This suggests that the loop affects the neighbouring vacancies along the normal direction of the loop's plane. However, at a higher temperature of 1000 K, i.e., Figure 5 (c) and (f), the MSD values for both systems are very similar with the MSD in the *x-y* plane is higher than that along the *z* direction. Further, as reported by [40,67,68], vacancies in α-Zr shows faster diffusion in the basal plane (*x-y* plane in this study) than along the *c*-axis (*z* direction). Thus, at higher temperatures the effect of the loop on the vacancies and clusters surrounding becomes much less important than the 'inherent' mobility of vacancies and their clusters, while at lower temperatures the loop effect is more significant.

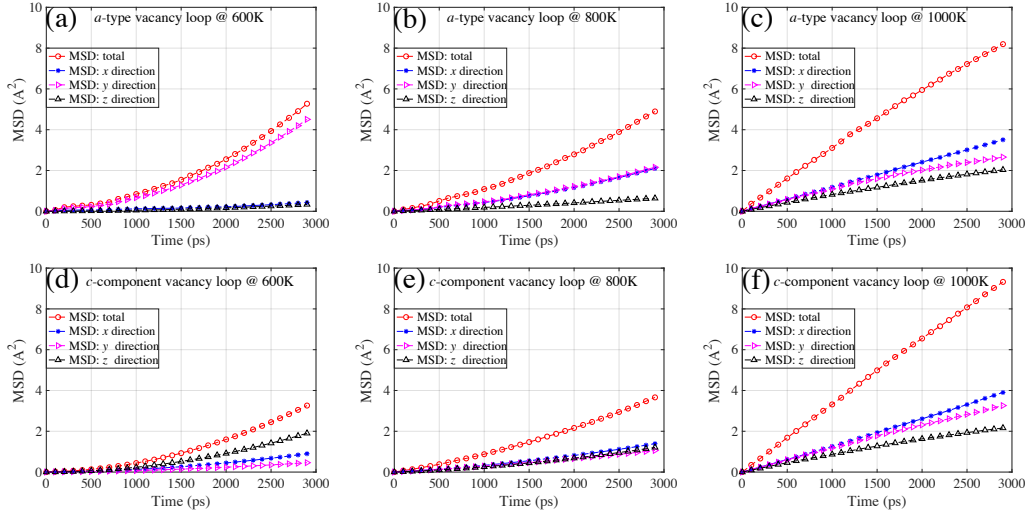

**Figure 5. The mean square displacement (MSD) in different directions during the growth of $a$-type and $c$-component vacancy loops at different temperatures, as a function of time.**

The vacancy loop growth mechanism in a vacancy supersaturated environment is further analyzed by determining the number of monovacancies $N_M$, divacancies $N_D$ and clusters $N_C$ during the loop growth, which is reported in Figure 6. Note that monovacancies are not considered as vacancy clusters, but that divacancies are. Both $N_M$ and $N_D$ for $a$-type vacancy loops at 600 K decrease during the first 1 ns, in tandem with loop growth as shown in Figure 4 (a). At the same temperature for $c$-component vacancy loops, only $N_D$ decreases while $N_M$ is relatively constant. This suggests that at 600 K, $a$-type vacancy loops absorb both monovacancies and divacancies, while $c$-component vacancy loops mainly absorb divacancies. As shown in the animations of the Supplementary Materials [64], at 600 K, $a$-type vacancy loops are more mobile than $c$-component vacancy loops, which allows $a$-type vacancy loops to glide to other positions and hence absorb monovacancies. At 1000 K, as illustrated in Figure 6 (d-f) the values of $N_M$, $N_D$ and $N_C$ for both loops continuously decrease, which is consistent with the loop growth behavior in Figure 4 (c). Thus, at such high temperature, the growth of $a$-type and $c$-component vacancy loops occurs by absorbing all kinds of vacancies and clusters. The ability of $a$-type vacancy loops to tilt and glide in such a way, increasing its probability of capturing defects, is typically not embedded in kMC and rate-theory models. Our results suggest that the capture radius of $a$-type loops in the prismatic direction should be larger than in the basal direction or than the capture radius of $c$-component loops of the same diameter.

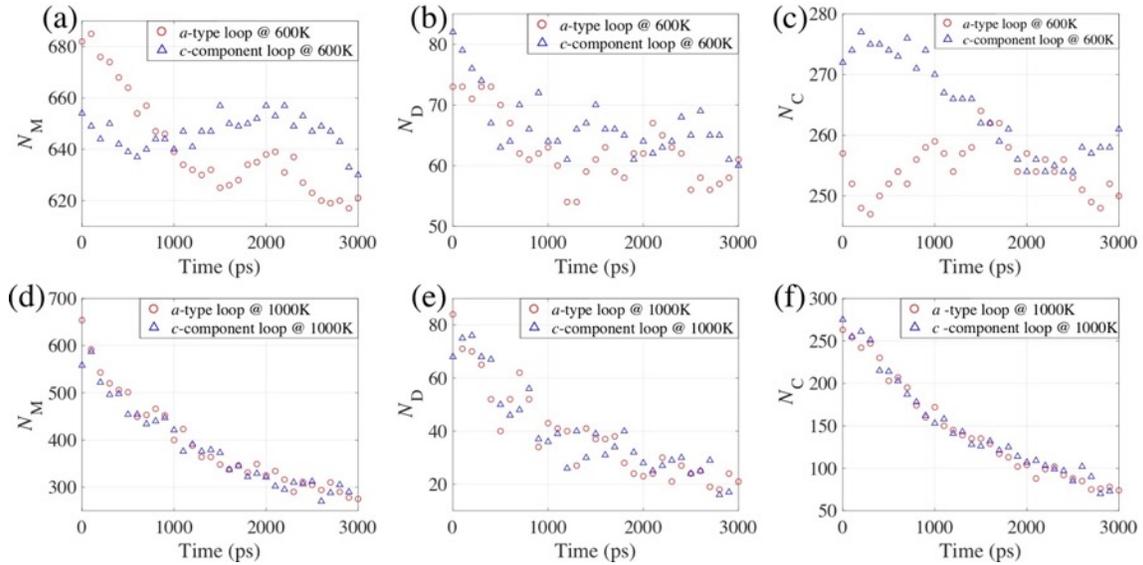

**Figure 6. The number of monovacancies as a function of time during the growth of a vacancy loop at (a) 600 K and (d) 1000 K; the number of divacancies as a function of time at (b) 600 K and (e) 1000 K; the number of total vacancy clusters as a function of time at (c) 600 K and (f) 1000 K.**

Visual inspection of the animation presented in the supplementary information [64] shows that the growth of both *a*-type and *c*-component dislocation loops is mainly due to the absorption of vacancies and vacancy clusters at the edge of the loops. We then calculated the formation energy of vacancies at different positions relative to the loop ($R$=40 Å) as a function of the radial and normal direction, which is shown in Figure 7, (b) for an *a*-type vacancy loop, and (c) for a *c*-component vacancy loop. There are clear minima of the potential energy of vacancies at the edge of the loops, which is consistent with the visual inspection of the animations. This result is also consistent with previous calculations of the strain and stress fields around dislocation loops and lines [75-77]. Similarly, vacancies preferentially form near grain boundaries in Zr [65] and in cubic materials [78,79].

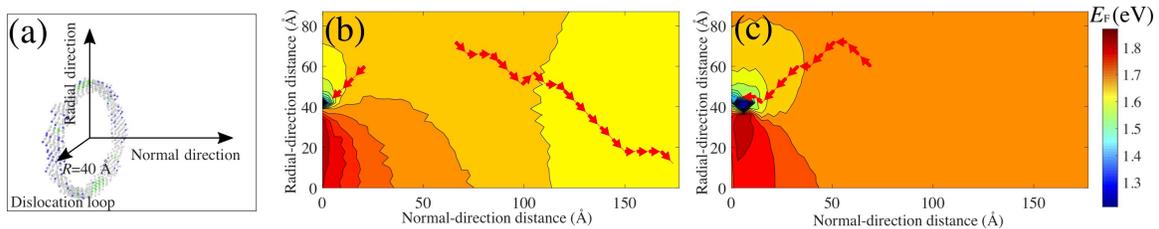

**Figure 7. (a) The normal direction and the radial direction relative to the dislocation loop. The formation energy of a vacancy as a function of the normal direction and the radial direction to the *a*-type vacancy loop (b) and *c*-component vacancy loop (c). In (b) and (c), the connected red dots indicate steepest-gradient paths. In (b) these paths indicate a driving force towards the edge of the loop and a driving force away from the loop, in its normal direction. In (c), the path indicates a driving force towards the edge of the loop. (For interpretation of the references to color in this figure legend, the reader is referred to the web version of this article.)**

As observed by Harte et al. [80] and in other experimental studies [5,81], *a*-type loops are randomly distributed at lower irradiation doses, and loops' separation distances are relatively large. Thus, the growth of a population of dislocation loops at low dose will be similar to a single loop growth. The density of *a*-type loops increases with dose. Moreover, these studies [5,80,81] also found that the majority of *a*-type loops at higher irradiation doses share a basal plane, which would decrease their separation distance. By observing the vacancy formation energy map in Figure 7, it seems that vacancies that follow an energy gradient will either converge towards the edge of the loop or be driven away in a direction perpendicular to the loop. Two such possible trajectories are illustrated in Figure 7 (b). Also, notice that the gradient pushing vacancies away from the loop in a direction normal to it, tends to push it in a position where it shares the loop's basal plane. These energy gradients may explain why multiple loops are experimentally found lying on the same basal plane, roughly equally spaced apart.

As a whole, our results show that properly modelling the temperature and strain-dependence of the diffusion coefficients of vacancies and vacancy-clusters is crucial in capturing different growth regimes at different temperatures. In the particular case of zirconium, this means that different relative population of *a*-type and *c*-component vacancy loops will be observed near GBs for a given dose, depending on irradiation temperature.

It should be noted that the MA07 potential (#3) selected in the present study is notorious for underestimating the binding energies of divacancies and trivacancies [40]. Since vacancy loop growth is affected by the number of neighbouring monovacancies, divacancies, and other small clusters, the vacancy loop growth simulations presented here are not quantitatively accurate. For example, the real transition from one growth regime to another will most likely *not* take place at 800K. Instead, it will take place at another temperature.

## 3.5 The stability of an $a$-type vacancy loop near the tilt GB during collision cascades

The interaction of an $a$-type vacancy loop with a 50 keV PKA near the GB was investigated in pure α-Zr. The collision cascade induced a rotation and gradual glide towards the GB, which led to eventual absorption by the GB as illustrated in Figure 8 (e).

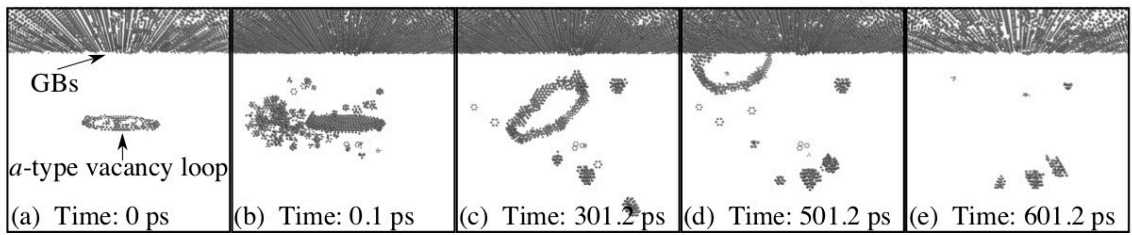

**Figure 8. Snapshots illustrating a cascade-induced absorption of an $a$-type vacancy dislocation loop by the GB at 573 K. The clock starts as the PKA is introduced. Only displaced atoms are shown based on the common neighbor analysis.**

The importance of Fe alloying on the formation of $c$-component loops was discussed by Carlan et al. [82], and a recent study by Topping et al. [83] suggested that an increase in Fe content in a Zr alloy delays $c$-loop nucleation. Here, the effect of Fe on the loop stability is explored. Collision cascades were performed near a pre-existing $a$-type vacancy loop with a 50 keV PKA in the vicinity of a tilt GB. The distance between the GB and the loop is about 10 nm, which is the same size as the diameter of the loop (see

Figure 9 (a)). The hybrid MD/MC method was applied to perform Fe segregation to the loop.

Fe segregation stabilized the loop, which did not undergo cascade-induced absorption, as illustrated in Figure 9. An animation of the cascade simulation is provided in the Supplementary Materials [64]. PKAs at multiple different locations and directions were simulated in this system, and all the results showed that the loop was stabilized by the segregated Fe atoms during the collision cascade.

Naturally, the strain field induced by the GBs gradually decreases as the separation distance between the loop and the GBs increases. One would expect the loop to have a decreased propensity to be absorbed by the GBs. However, this relation has not been systematically studied. In addition, different types of GBs may produce different levels of strain field that would then affect their sink strength as suggested in [32,65]. In the future, studying these factors would be of interest.

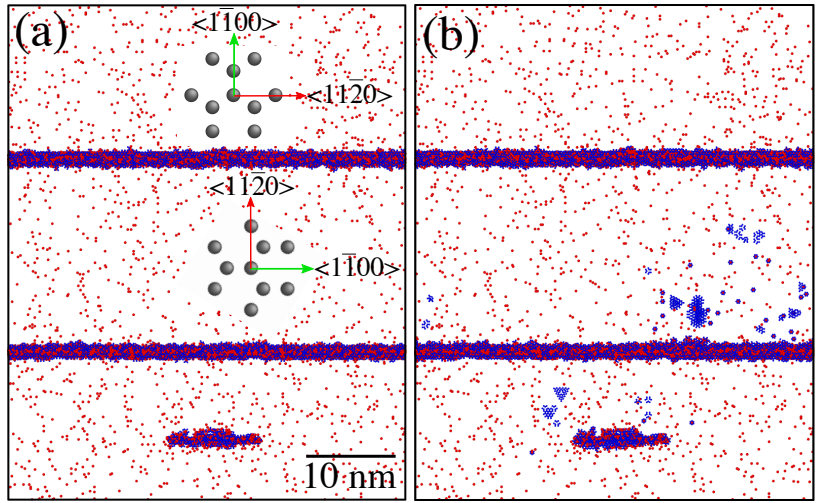

**Figure 9. Snapshots (a) before and (b) after a 50 keV collision cascade overlapping with an Fe-enriched *a*-type vacancy loop at 573 K. Fe atoms are colored in red, and the displaced atoms are shown here based on common neighbor analysis.**

To understand how alloying Fe can stabilize the dislocation loop, the effect of Fe segregation on the loop's stress field was explored. Five nm radius interstitial and vacancy loops of *a*-type and *c*-component were equilibrated at 573 K. Figure 10 illustrates the Fe concentration in the plane of the dislocation loop after the relaxation, and Fe segregation was greatest at interstitial loops. Fe atoms segregated to the inside of the *a*-type interstitial loop, as illustrated in Figure 10 (a). The concentration of Fe in the center of the loop is expected to decrease as loop size increases [76]. Figure 10 (d) shows that the *c*-component vacancy loop exhibits the least Fe segregation.

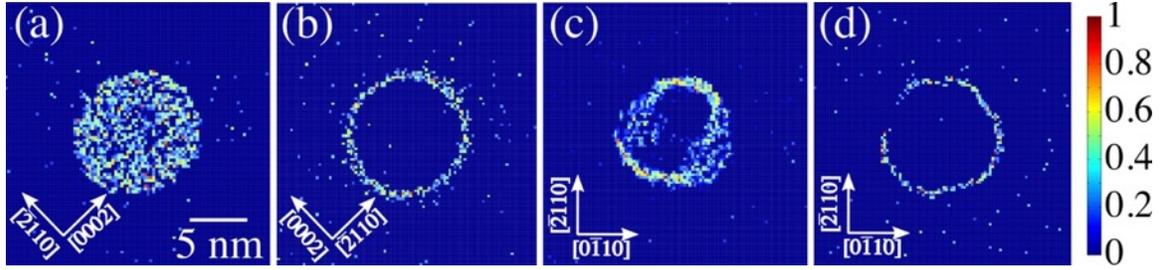

**Figure 10. The Fe concentration map of dislocation loops in the Zr-Fe system: (a) *a*-type interstitial loop, (b) *a*-type vacancy loop, (c) *c*-component interstitial loop, (d) *c*-component vacancy loop. (For interpretation of the references to color in this figure legend, the reader is referred to the web version of this article.)**

The distortion of the dislocation loop (elastic energy) and segregation of Fe atoms (chemical energy) results in a change of the system energy. Defect energy ($E_{Def}$) is defined the energy variation of an atom due to interaction with the dislocation loop or chemical segregation: $E_{Def}=E_{Pot}-E_{Per}$, where $E_{Pot}$ is the potential energy of any atom, and $E_{Per}$, is the averaged potential energy of an atom in a perfect Zr lattice at equilibrium. The defect energy maps in the loop's plane without Fe segregation are shown in Figure 11 (a-d), and they are *a*-type interstitial loop, *a*-type vacancy loop, *c*-component interstitial loop and *c*-component vacancy loop, respectively. Figure 11 (e-h) are the defect energy maps corresponding to Figure 11 (a-d), which are in the condition of Fe segregation. By comparing Figure 11 (a-d) and Figure 11 (e-h), we can see that the values of $E_{Def}$ are greatly reduced by Fe segregation to the loops, which explains why loops become more stable after Fe segregation.

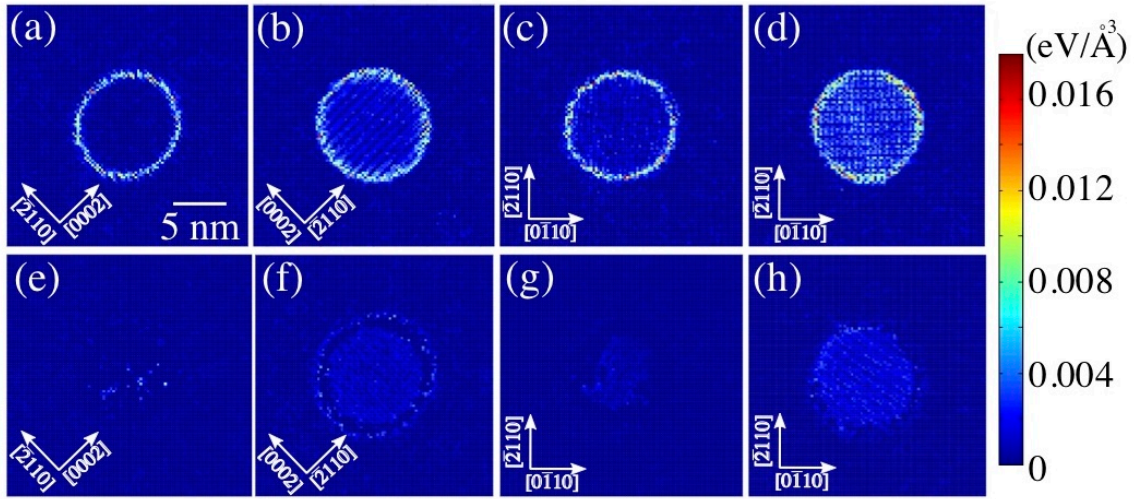

**Figure 11.** The defect energy map of dislocation loops either without Fe segregation (a-d) or with Fe segregation (e-f): (a) and (e) are $a$-type interstitial loop, (b) and (f) are $a$-type vacancy loop, (c) and (g) are $c$-component interstitial loop, (d) and (h) are $c$-component vacancy loop. (For interpretation of the references to color in this figure legend, the reader is referred to the web version of this article.)

The radial-direction component of atomic stresses ($S_{rr}$) of dislocation loops with and without Fe segregation were calculated in Figure 12. The $S_{rr}$ of interstitial loops is greatly affected by Fe segregation. Without Fe segregation, as illustrated in Figure 12 (a) and (c), $S_{rr}$ is compressive inside the loop. However, $S_{rr}$ became tensile after Fe segregation to the loops, see Figure 12 (e) and (g). Moreover, the magnitude of the contrast of $S_{rr}$ near the edge of the loop is dramatically decreased after Fe segregation for all types of the loop.

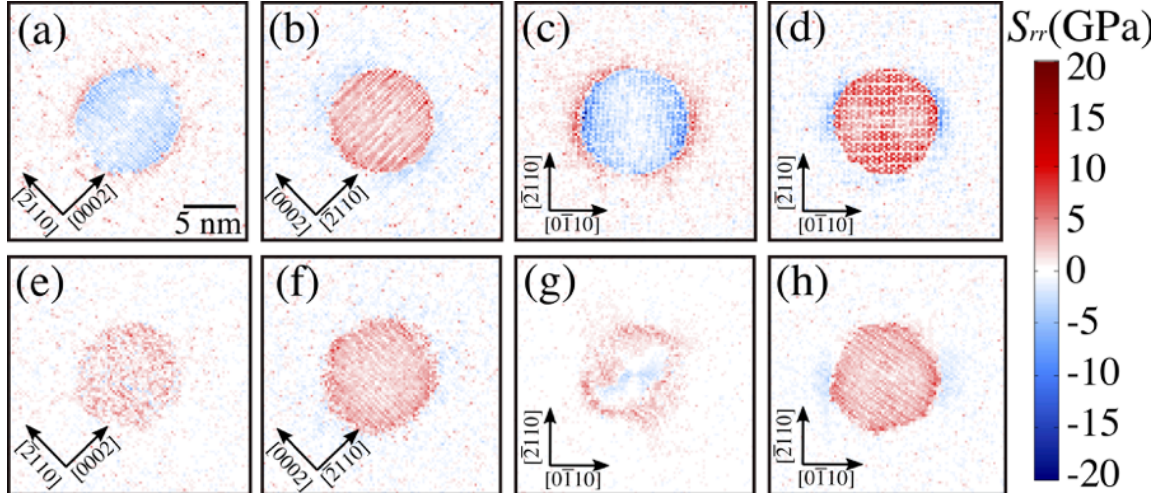

Figure 12. The radial-direction component of stress ($S_{rr}$) for dislocation loops either without Fe segregation (a-d) or with Fe segregation (e-f): (a) and (e) are *a*-type interstitial loop, (b) and (f) are *a*-type vacancy loop, (c) and (g) are *c*-component interstitial loop, (d) and (h) are *c*-component vacancy loop. Red is tensile and blue is compressive. (For interpretation of the references to color in this figure legend, the reader is referred to the web version of this article.)

Recently, Harte et al. [80] hypothesized that a row of *a*-type loops that are aligned parallel to the trace of the basal plane might collapse into a coarse *c*-component loop. Fe segregation can stabilize *a*-type loops as evidenced by the results above, which may impede the transition from the aligned *a*-type loops to a coarse *c*-component loop. This may explain why *c*-component loop nucleation is delayed in Zr-Fe alloys, in comparison to pure Zr, as reported by Topping *et al.* [83]. In addition, the enhanced stabilization of *a*-type loops by the presence of Fe could prevent the absorption of *a*-type loops by the GBs. This may affect the number of free vacancies and their clusters near the GBs due to the growth of *a*-type loops.

## 4 Conclusion

Primary cascade production of various defects has been statistically studied. The defect production results in the current study provide valuable additions to databases of production rates which are necessary for kMC and rate theory models in the context of predicting radiation damage build-up. The following conclusions were drawn:

1. Fewer residual defects remain after collision cascades in the presence of extended defects, compared to a perfect crystal. In particular, this will contribute to vacancy supersaturation near the GBs, which could promote vacancy loop growth.

2. $a$-type vacancy loops are more stable than $a$-type interstitial loops near the GBs. Also, simple diffusion sink arguments predict a surplus of vacancy-type defects near GBs. The hypothesis of vacancy supersaturation near GBs is further supported by the results of *in situ* TEM experiments in this study, which found a high density of vacancy loops near the GBs.

3. Vacancy loops have been simulated in a vacancy supersaturated environment. Growth of $a$-type vacancy loops is faster than that of $c$-component loop at low temperatures (e.g., 600 K), but both types of loops have similar growth rates at high temperatures (e.g., 1000 K) since the diffusion of vacancies and their clusters are dominant at high temperatures. Thus, the ratio of $a$-to-$c$ loops observed near GBs is expected to be dependent on temperature. Also, the simulations suggests that the capture radius of $a$-type vacancy loops in the prismatic direction should be larger than that in the basal direction and that those of $c$-component loops. Furthermore, analyzing the energy landscape around an $a$- type vacancy loop also gave insights into why they are typically observed lying on common basal planes, with somewhat regular inter-loop spacing.

4. $a$-type vacancy loops can be absorbed by GBs during cascade-induced events.

5. Fe segregation to the loops enhances loop stability by reducing their defect energies and the magnitude of the stress field at the edge of the loops. This effect impedes cascade-induced absorption of loops by the GBs.

**Acknowledgements**

The authors thank Compute Canada for generous allocation of computer resources. The research was supported by the NSERC and the NSERC/UNENE Industrial Research Chair in Nuclear Materials at Queen's.